\documentclass[11pt,a4paper]{article}
\usepackage[utf8]{inputenc}
\usepackage[T1]{fontenc}
\usepackage{lmodern}
\usepackage[english]{babel}
\usepackage[margin=2.5cm]{geometry} 
\usepackage{graphicx} 
\usepackage{booktabs} 
\usepackage[round,authoryear]{natbib}
\usepackage{amsmath} 
\usepackage{amssymb} 
\usepackage{setspace} 
\usepackage{caption} 
\usepackage[colorlinks=true, citecolor=blue, linkcolor=blue, urlcolor=blue]{hyperref} 
\usepackage{xltabular}
\usepackage{array}

\usepackage{bbm}
\usepackage{mathrsfs}
\usepackage{amsfonts}
\usepackage{float}
\usepackage{amsthm}
\usepackage{epsfig,rotating}
\usepackage{amssymb,amsmath}
\usepackage{latexsym}
\usepackage{appendix}
\usepackage{graphicx,subfigure}
\usepackage{multirow}
\usepackage{color,url}
\usepackage{booktabs}
\usepackage[ruled,linesnumbered]{algorithm2e}
\usepackage{subfigure}
\usepackage{float}
\usepackage{comment}

\usepackage{ragged2e}
\usepackage{titlesec}
\usepackage{xurl}
\newcolumntype{Y}{>{\RaggedRight\arraybackslash}X}
\newcolumntype{L}[1]{>{\RaggedRight\arraybackslash}p{#1}}
\newcolumntype{C}[1]{>{\centering\arraybackslash}p{#1}}

\titleformat{\section}[hang]{\normalfont\Large\bfseries}{\thesection}{1em}{}
\titleformat{\subsection}[hang]{\normalfont\large\bfseries}{\thesubsection}{1em}{}

\def\b1{\boldsymbol{1}}
\def\eps{\varepsilon}

\def\spacingset#1{\renewcommand{\baselinestretch}%
{#1}\small\normalsize} \spacingset{1}

\usepackage{amsmath}
\usepackage{tikz}
\bibpunct{(}{)}{;}{a}{,}{,}

\renewcommand{\footnotesize}{\fontsize{9}{11}\selectfont}

\title{Understanding Sectoral Responses to Tariff Uncertainty
via a High-Dimensional Mediation Analysis\footnote{Qilan Hong is a senior high school student
at Polytechnic School, Pasadena, CA 91106, and Runze Li is Eberly Family Chair Professor of
Statistics, The Pennsylvania State University, University Park, PA 16802. The authors would like to thank
Dr. Ye Yu for her constructive suggestions and helpful discussions on real data analysis.}
}
\author{
    Qilan Hong and Runze Li
}
\date{} 

\begin{document}

\maketitle

\begin{abstract}
This paper conducts an empirical mediation analysis  to examine how U.S. stock-market sectors responded to the 2025 tariff-policy episode. Using S\&P 500 stock data along with hundreds of firm-level financial variables, we study whether sector membership was associated with differences in stock returns and whether these differences were statistically related to firm financial characteristics. For each of five economically distinct tariff-policy windows, we use high-dimensional linear causal mediation models to estimate the direct sector effects. It is noteworthy that estimating the indirect section effects is challenging using the linear causal mediation models since the indirect sector effect is the product of a high-dimensional regression coefficients and a high-dimensional coefficient matrix.
Instead, we use a linear regression model to estimate the total sector effects, and then estimate the indrect sector effects by using the difference between the total sector effects and the direct sector effects.
To accommodate a large number of candidate mediators, we use a partially penalized least-squares procedure that regularizes financial-variable coefficients while unpenalizes coefficients of all sector indicators. Wald and F-type tests are used to examine whether indirect and direct effects are significant or not, respectively. 
The results show that tariff news did not produce one uniform market response. Sector effects vary across 
five economically distinct tariff-policy windows.
The clearest findings are in the direct sector effects. Technology was the sector most negatively affected during the initial decline and escalation collapse, but it also rebounded most strongly during the April 9 relief rally.  
The results indicate that in addition to Technology, Energy and other trade-exposed sectors were particularly affected during the initial decline and escalation phases.
The selected mediators  provide evidence that growth, investment, liquidity, profitability, and cash-flow characteristics helped characterize firms associated with sector-level return differences. 
\end{abstract}


\textbf{Keywords:} Lasso, Linear mediation models, Partial penalized least squares, SCAD.

\textbf{Mathematics Subject Classification (2020):} 62F03, 62J07

\textbf{JEL classification}: C12, C13

\baselineskip=24pt


\newpage

\section{Introduction}\label{sec}

Tariff policies can pose significant effects for firms, such as
increasing the cost of imported materials or uncertainty about
future trade policies. Though, such effects vary by company. Some
sectors depend on international trade more than others; firms in
the same sector also can differ in metrics like profitability.
This motivates us to study how financial market reacts to a new
tariff policy.

The United States in 2025 was a highly volatile period in its
constant tariff and policy shifts, providing a useful opportunity
to examine these differences. In the stock market, multiple major
movements were motivated by tariff announcements, pauses,
retaliations, and escalations. Due to the complexity of this
period, only looking at the overall market returns may therefore
overlook imperative differences in how individual firms responded.
Previous research has also found that aspects like prices,
investment decisions, and economic activity are all affected by
tariffs and the uncertainty surrounding trade policy
\citep{amiti_redding_weinstein_2019,handley_limao_2015,caldara_et_al_2020}.
Thus, it is of great interest to study how financial market reacts
differently to a new tariff policy using 2025 stock data.

In this paper, we aim to examine the potential causal effects of
sector exposure. Specifically, we target to address two important
questions: (1) whether the impact of sector membership on stock
return are different over different tariff-policy windows, and (2)
whether those differences were associated with firms' financial
characteristics. To this end, we collect S\&P 500 data from
internet, and construct 215 potential financial variables, which
leads us to consider high-dimensional linear mediation framework
in our empirical analysis. Hence we apply the statistical
inference procedures for high-dimensional linear mediation models
proposed in \citet{guo2023highdim} in our data analysis, in which
sector membership is treated as the exposure variable, firm's
financial metrics are taken as potential mediators and stock
log-return is set to the outcome. See more detailed definitions of
these variables in Section 2.


During 2025, there were several tariff episodes. A single return
representative of the full 2025 period cannot fully represent
these different phases. This motivates us to consider five tariff
windows, comparing the short-window results with a longer
adjustment window. Table~\ref{tab:event-windows} presents the
exact dates of the five tariff windows and rationales on how to
define these windows. For each windows, we carry out an empirical
analysis and then examine the direct effect and indirect effect of
sector exposure on stock returns.

Our empirical analysis indicates that the impacts of sector
membership on stock return can be quite different over different
periods. For example, both direct effect and indirect effect of
sector exposure are significant over the initial tariff-driven
decline period, but neither direct effect nor indirect effect is
significant over the long-run adjustment. Our empirical results
implies that that the indirect effect of sector exposure is
significant at level 0.05 only for the initial period. This
implies the sector exposure has significant impact on the stock
return through financial metrics only during the initial period.
This might be interpreted as during the initial period, financial
metrics for most stock are highly associated with sector exposure,
but after the initial period, most firms in S\&P 500 got well
prepared for tariff, and the financial metrics become insensitive
to sector exposure. Our empirical results further show that sector
differences changed substantially across the five windows, and the
selected financial variables also varied across scenarios.
Relationships between the variables vary across the five windows.

The rest of this paper is organized as follows. Section 2 presents
research problems, data sources and how to construct features and
response. Section 3 presents the statistical model, estimation
procedure and statistical inference procedure used in this paper.
Section 4 is devoted to detailed empirical analysis. Conclusions
are presented in Section 5. Variable dictionary, additional
summaries and numerical results are presented in the Appendix.

\section{Research Problems, Data Sources and Feature and Response Constructions}
\label{sec:data-problem}

\subsection{Research problem}
\label{subsec:research-problem}

It is of great interest to study how financial market reacts to a
new economic policy such as a new tariff policy. In general,
tariffs did not create a uniform market shock. Firms had different
circumstances in their exposure to factors such as imported
inputs, export demand, global supply chains, commodity prices, and
financing conditions. Thus, market responses may vary by sector
and by the financial structure of firms in such sectors. This
motivates us to investigate that, during the 2025 tariff-policy
period, (a) how S\&P 500 stock returns differed across sectors,
and (b) whether those differences were related to firms' financial
characteristics.

During 2025, there were several tariff episodes. Specifically, on
February 1, the administration announced additional tariffs on
imports from Canada, Mexico, and China, including 25 percent
tariffs on Canada and Mexico, a 10 percent tariff on China, and a
lower 10 percent rate on Canadian energy
\citep{whitehouse_tariffs_canada_mexico_china_2025}. On March 4,
these tariffs took effect and prompted retaliation from major
trading partners \citep{ap_tariffs_retaliation_2025}. On March 12,
expanded steel and aluminum tariff actions became effective,
widening the tariff episode from country-specific measures to
direct input-cost pressure on industrial supply chains
\citep{whitehouse_aluminum_2025, whitehouse_steel_2025}. the
market initially fell during the February to March tariff
escalation, and dropped in an even sharper manner after the
reciprocal tariff announcements in late March and early April. It
then rebounded following the April 9 tariff pause. Later, it
declined again as tensions with China, concerns over
pharmaceutical tariffs, and uncertainty about monetary policy
increased. A single return representative of the full 2025 period
would not accurately represent these different phases. This
motivates us to consider five tariff windows, comparing the
short-window results with a longer adjustment window.
Table~\ref{tab:event-windows} presents the exact dates of the five
tariff windows and rationales for defining these windows.

It is well known that individual stock return is highly associated
with its financial metrics. Thus, it is also of great interest to
study whether financial metrics can help mediate the connection
between
the company sector and the return on investment during each tariff window. 
To address research problems (a) and (b) and understand the
role of financial metrics under different tariff scenarios, we
will focus on two aspects: (1) to examine which sectors
outperformed or underperformed the referenced sector: the
utilities sector, and (2) to check if any sector-return differences can be
statistically associated with accounting characteristics such as
profitability, liquidity, debt, investment, operating costs, or
cash-flow strength.

\begingroup
\small \setlength{\tabcolsep}{3pt}
\renewcommand{\arraystretch}{1.15}

\begin{xltabular}{\textwidth}{L{1.4cm} L{3.0cm} L{3.5cm} Y}
\caption{Tariff-policy return windows}
\label{tab:event-windows} \\
\toprule
Scen. & Study label & Close-to-close window & Rationale \\
\midrule
\endfirsthead

\multicolumn{4}{l}{\textit{Table \thetable{} continued}} \\
\toprule
Scen. & Study label & Close-to-close window & Rationale \\
\midrule
\endhead

\bottomrule
\endlastfoot

S1 & Initial tariff-driven decline & February 19 close--March 13 close, 2025 & Starts at the S\&P 500 local peak around February 19 and ends at the March 13 local low. This window captures the initial selloff after the February fentanyl-related tariff orders, the March 4 reinstatement/escalation of Mexico, Canada, and China tariffs, and the March 12 steel and aluminum tariffs. \\

S2 & Escalation collapse & March 25 close--April 8 close, 2025 & Starts after the partial recovery into March 25 and ends at the April 8 trough. This window captures the Venezuela-oil tariff order, the April 2 ``Liberation Day'' reciprocal-tariff announcement, China's retaliation, and the April 7 threat to add further tariffs on China. \\

S3 & Policy-shock relief jump & April 8 close--April 9 close, 2025 & Captures the sharp rebound from the April 8 trough to the April 9 jump after the 90-day pause on reciprocal tariffs for non-retaliating countries, alongside the increase in China's tariff rate. \\

S4 & Uncertainty-driven decline & April 9 close--April 21 close, 2025 & Starts after the relief jump and ends at the April 21 decline. This window captures renewed uncertainty from China-related escalation, pharmaceutical-tariff concerns, and public pressure on Federal Reserve Chair Jerome Powell. \\

S5 & Long-term adjustment & February 19 close--December 31 close, 2025 & Measures the broader adjustment from the February peak through the rest of 2025, allowing the short-run tariff shocks, subsequent recoveries, and later stabilization to appear in one long-window outcome. \\

\end{xltabular}

\endgroup

\subsection{Data sources}
\label{subsec:data-sources}

We collected S\&P 500 stock data (indeed 503 stocks) using a
frozen universe. The universe was obtained through Financial
Modeling Prep's (FMP) S\&P 500 constituent service and then stored
as a fixed stock list for every analysis stage. S\&P Dow Jones
Indices describes the S\&P 500 as a float-adjusted
market-capitalization-weighted index of leading large-cap U.S.
companies \citep{spglobal_sp500}.

All market, company-profile, and accounting inputs were obtained
under an FMP Plus subscription through FMP's \texttt{stable} API.
The final verified collection was performed on June 27--28, 2026.
Quarterly accounting data came from the \texttt{income-statement},
\texttt{balance-sheet-statement}, and \texttt{cash-flow-statement}
endpoints \citep{fmp_stable_api}. Sector labels came from the
\texttt{sector} field of the FMP company-profile endpoint
\citep{fmp_company_profile}. Daily prices came from FMP's
dividend-adjusted historical price endpoint, which supplies
end-of-day prices adjusted for dividend distributions
\citep{fmp_dividend_adjusted_prices}.

The dataset was created in five steps. First, the S\&P 500 stock
list was frozen. Second, the pipeline collected quarterly
income-statement, balance-sheet, and cash-flow data for each
stock. Third, each stock was assigned one sector from the FMP
company-profile data. Fourth, daily dividend-adjusted prices were
collected for 2025. Fifth, the daily returns were split into five
tariff-policy windows chosen from the S\&P 500 turning points and
the policy timeline documented in the scenario notes.

The validated dataset contains 503 stocks and 43 raw quarterly accounting fields. Because index membership and vendor data can be
revised, the frozen universe, retrieval dates, endpoint names, and
code version accompanies the archived article materials.  As shown
in Table~\ref{tab:event-windows} along with further details in
Table~\ref{tab:appendix-return-coverage}, the five windows
separate the tariff episode into economically different market
phases.

\subsection{Feature and Response Constructions}

\subsubsection{Outcome construction}
\label{subsec:outcome-construction}

The response variable was constructed as follows. We first
collected FMP dividend-adjusted daily closing prices. For each
stock and each trading day, the code first computes a daily log
return. A log return is used because daily log returns can be
added across a window to create one multi-day return measure. This
is how the code turns a sequence of daily stock prices into one
return number for each tariff window.

Define $P^{\mathrm{adj}}_{i,t}$ to be the dividend-adjusted
closing price of stock $i$ on trading date $t$. The daily log
return is
\begin{equation}
r_{i,t}=\log\left(\frac{P^{\mathrm{adj}}_{i,t}}{P^{\mathrm{adj}}_{i,t-1}}\right),
\end{equation}
and the outcome for window $w$ is the sum of those daily log returns:
\begin{equation}
Y_{i,w}=\sum_{t\in w} r_{i,t}.
\end{equation}

\subsubsection{Sector exposure} \label{subsec:sector-exposure}

Each stock was assigned the sector according to its sector
membership reported by FMP's company-profile endpoint and encoded
as one of eleven mutually exclusive indicator variables.
Table~\ref{tab:appendix-sector-counts} depicts the number of
stocks in each sector.

In our empirical analysis in Section 4, the sector of utilities is
set to be the reference group. 
Thus, every reported sector coefficient is a difference relative
to Utilities. The illustrative companies below are recognizable
members of the frozen study universe. It is noteworthy that they
are examples only and were not used to determine the definitions.

\begingroup
\small
\setlength{\tabcolsep}{4pt}
\renewcommand{\arraystretch}{1.08}

\begin{xltabular}{\textwidth}{L{3.0cm} Y L{4.0cm}}
\caption{Sector definitions and illustrative companies}
\label{tab:sector-definitions} \\
\toprule
Sector & Sector definition & Illustrative companies \\
\midrule
\endfirsthead

\multicolumn{3}{l}{\textit{Table \thetable{} continued}} \\
\toprule
Sector & Sector definition & Illustrative companies \\
\midrule
\endhead

\midrule
\multicolumn{3}{r}{\textit{Continued on next page}} \\
\endfoot

\bottomrule
\endlastfoot

Basic Materials &
Producers and processors of chemicals, industrial gases, construction materials, metals, mining products, paper, and related raw materials. &
Linde; Newmont; Freeport-McMoRan \\

Communication Services &
Providers of telecommunications, wireless and network services, media, entertainment, interactive media, and digital communication platforms. &
Alphabet; Meta Platforms; Verizon \\

Consumer Cyclical &
Businesses whose demand tends to vary with household income and the economic cycle, including autos, retail, consumer durables, leisure, restaurants, and travel. &
Amazon; Tesla; Home Depot \\

Consumer \par Defensive & Producers and retailers of frequently
purchased necessities, including food, beverages, household
products, personal products, and staples distribution. &
Walmart; Costco; Coca-Cola \\

Energy &
Companies involved in oil, gas, and consumable-fuel exploration, production, refining, transportation, storage, and related equipment and services. &
Exxon Mobil; Chevron; ConocoPhillips \\

Financial Services &
Banks, insurers, capital-markets firms, consumer-finance providers, payment networks, asset managers, and diversified financial companies. &
Berkshire Hathaway; JPMorgan Chase; Visa \\

Healthcare &
Drug and biotechnology developers, healthcare providers, medical-technology and equipment manufacturers, life-sciences businesses, and healthcare-service companies. &
Eli Lilly; Johnson \& Johnson; AbbVie \\

Industrials &
Manufacturers and service providers in capital goods, aerospace and defense, machinery, transportation, construction, professional services, and commercial support activities. &
Caterpillar; GE Aerospace; RTX \\

Real Estate &
Equity real-estate investment trusts and companies that own, operate, develop, or manage income-producing property and related real-estate services. &
Welltower; Prologis; Equinix \\

Technology &
Producers of software, semiconductors, computing hardware, electronic equipment, and information-technology services. &
NVIDIA; Apple; Microsoft \\

Utilities &
Providers of regulated or merchant electricity, gas, water, multi-utility, and independent power services. This is the model's reference sector. &
NextEra Energy; Southern Company; Duke Energy \\

\end{xltabular}
\endgroup

\subsubsection{Financial variables and transformations}
\label{subsec:financial-variables}

For every stock, the pipeline obtained up to 40 quarters of
income-statement, balance-sheet, and cash-flow data from FMP. For
each statement, the analysis identified an early-2025 anchor
quarter: the latest available statement period dated between
January 1 and April 30, 2025. The anchor quarter is not uniformly
calendar-quarter Q1 because firms follow different fiscal
calendars. Most firms have an anchor date of March 31, 2025, while
the selected dates range from January 31 to April 30, 2025.

The final dataset contains 17 raw income-statement variables, 22
raw balance-sheet variables,  and four raw cash-flow variables.
Each raw series $v$ was expanded into five candidate features. Let
$v_{i0}$ denote the value for firm (i) in the selected early-2025
anchor quarter, $v_{i1}$ the value from the immediately preceding
quarter, and $v_{i4}$ and $v_{i12}$ the values four and twelve
quarters before the anchor quarter. The transformations are:

These transformations yield 215 financial metrics that may be
considered for mediators. Before estimation, all-missing and
zero-variance mediators were removed. The validated dataset
contained one all-missing mediator, leaving 214 variables to be
standardized to mean zero and unit variance before our empirical
analysis. Remaining missing values were set to zero after
standardization. Definitions and preprocessing details are
reported in Section A.1 in the appendix.

\begin{xltabular}{\textwidth}{L{2.0cm} L{5.7cm} Y}
\caption{Financial-variable transformations}
\label{tab:financial-transformations} \\
\toprule
Suffix & Construction & Interpretation \\
\midrule
\endfirsthead

\multicolumn{3}{l}{\textit{Table \thetable{} continued}} \\
\toprule
Suffix & Construction & Interpretation \\
\midrule
\endhead

\bottomrule
\endlastfoot

none & $v_{i0}$ & Value in the selected early-2025 reference quarter \\
\texttt{\_YoY} & $(v_{i0}-v_{i4})/|v_{i4}|$ & Proportional change from the corresponding quarter one year earlier \\
\texttt{\_QoQ} & $(v_{i0}-v_{i1})/|v_{i1}|$ & Proportional change from the preceding quarter \\
\texttt{\_CAGR3} & $(v_{i0}/v_{i12})^{1/3}-1$ & Three-year compound annual growth; defined only when both endpoints are positive \\
\texttt{\_Avg5} & Mean of the selected reference quarter and up to 19 preceding quarterly values & Trailing average over at most 20 quarters \\

\end{xltabular}


\section{Statistical Models and Inference Procedures}
\label{sec:method}


\subsection{Linear mediation models}

To examine the impact sector membership, financial characteristics
on stock returns, it is natural to take the stock log-return as
the response $Y$, the sector indicators as the exposure variable
$X$, and the standardized financial characteristics as potential
mediators $M$, and consider a linear mediation model consisting of
two linear regression models
\begin{equation}
Y=\alpha_0^T M+\alpha_1^T X+\eps, \label{eq3-1}
\end{equation}
and
\begin{equation}
M=\Gamma^T X+\eta, \label{eq3-2}
\end{equation}
where $\eps$ is a random error with mean zero and variance
$\sigma^2_{\eps}$, and $\eta$ is a random error vector with mean
zero and covariance matrix $\Sigma_\eta$. To include an
intercept term in models (\ref{eq3-1}) and (\ref{eq3-2}), we set
the first element of $X$ to be 1.

Plugging (\ref{eq3-2}) into (\ref{eq3-1}), it follows that
\begin{equation}
Y=\alpha_0^T (\Gamma^T X+\eta)+\alpha_1^T X+\eps=(\Gamma\alpha_0 +
\alpha_1)^TX + (\alpha_0^T\eta +\eps)=\gamma^TX + \eps^*,
\label{eq3-3}
\end{equation}
where $\gamma= \beta + \alpha_1$ with $\beta=\Gamma\alpha_0$, and
$\eps^*=\alpha_0^T\eta +\eps$, which can be viewed as a random
error with mean zero.

\begin{figure}[!]
\usetikzlibrary{arrows.meta,positioning}
\begin{center}
\begin{tikzpicture}[
    >=Stealth,
    node distance=4cm,
    box/.style={
        draw=blue!70!black,
        thick,
        minimum width=5cm,
        minimum height=1.2cm,
        align=center
    }
]

\node[box] (X) at (0,0) {{\color{blue} X: Sector Membership}};
\node[box] (M) at (4,4) {{\color{blue} M: Financial Metrics}};
\node[box] (Y) at (8,0) {{\color{blue} Y: Stock Log-Return}};



\draw[->,blue!80!black] (0.2,0.7) -- (3.5,3.2);

\draw[->,blue!60!black] (4.5,3.2) -- (8.0,0.7);
\draw[->,blue!60!black] (2.6,0.0) -- (5.4,0.0);
\node at (1.6,2.0) {$\Gamma$};
\node at (6.4,2.0) {$\alpha_0$};
\node at (4.0,0.2) {$\alpha_1$};

\end{tikzpicture}
\end{center}
\caption{Diagram of Mediation Model}
\label{mediation}
\end{figure}
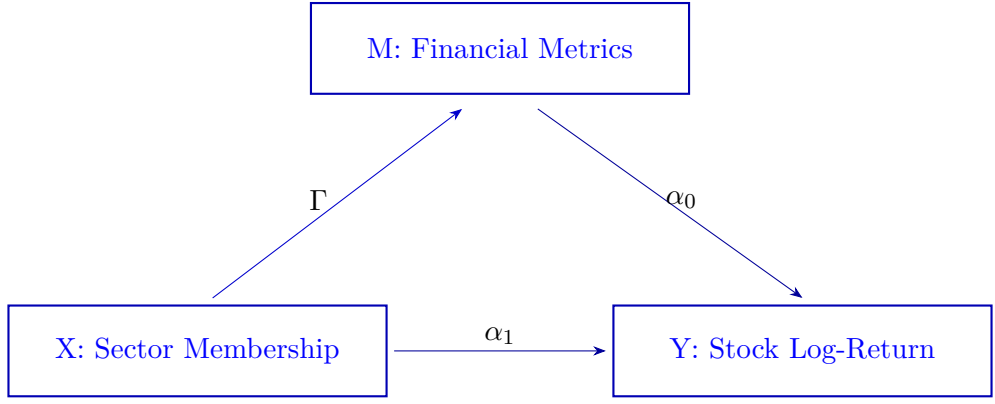

It is well known that the sector membership may result in
different financial metrics, and subsequently affect stock
returns. Figure~\ref{mediation} illustrates such a relationship among
the sector membership, financial metrics and stock log-return. Thus,
models (\ref{eq3-1}) and (\ref{eq3-2}) are particularly
useful to investigate the mediation effects of financial metrics
that bridge section membership and stock return.

In the literature of causal mediation analysis, (\ref{eq3-1}) is
referred to as outcome model regressing $Y$ on both exposure $X$
and mediators $M$, (\ref{eq3-2}) as mediator model regressing the
mediators on exposure variable, and (\ref{eq3-3}) as total effect
model.  In the literature, $\gamma$, $\alpha_1$ and $\beta$ are
termed as the total effect, direct effect and indirect effect of
exposure $X$, respectively. Under the independence conditions of
random errors in the models, and the sequential ignorability
assumption \citep{imai2010general}, $\alpha_1$ can be interpreted
as the average natural direct effect and $\beta$ can be
interpreted as the average natural indirect effects, of a one-unit
change in the exposure $X$ under the potential outcome model
framework (See, for example, Section S.2 in supplementary material
of \cite{guo2023highdim}).

\subsection{Estimation and inference procedure}
\label{subsec:estimation}

Suppose that for each tariff-policy window  $w$ and for each
stock, we collect data $\{Y_{iw}, X_{iw}, M_{iw}\}$, $i=1,\cdots,
n_w$. As stated in Table~\ref{tab:appendix-dataset-summary}, For
$w=1,\cdots, 4$, $n_w=503$, while $n_5=500$. In general, $X_{iw}$
does not vary over $w$ if the sector membership is invariant over
time. For ease of presentation, we will suppress the subscript $w$
since our empirical analysis will be carried out for each
tariff-policy winder separately. Thus, we simply denote $Y_i$ to
be the stock return of stock $i$, $X_i$ to be its sector
indicators, and $M_i$ to be its standardized financial
characteristics. Thus, data were collected from the linear
mediation model:
\begin{equation}
Y_i=\alpha_0^T M_i+\alpha_1^T X_i+\eps_{i}, 
\end{equation}
and
\begin{equation}
M_i=\Gamma^T X_i+\eta_i, \label{eq3-4}
\end{equation}
for $i=1,\cdots, n$.

As stated in Section 2.3.3, there are 214 financial metrics
available for our empirical analysis, however, only a relatively
small number of the possible  financial metric variables are
important for explaining stock returns. This motivates us to apply
penalized least squares to exclude irrelevant financial metric
variables. Since the sector effects is of primary interest, one
should not penalize the coefficients of $X$. Thus, the partial
penalized least squares estimation procedure developed in
\citet{guo2023highdim} is well suitable for our purpose.



The partially penalized least-squares function is defined as
\begin{equation}
 \frac{1}{2n} \sum_{i=1}^n (Y_i-M_i^T\alpha_0-X_i^T\alpha_1)^2 +
\sum_{j=1}^{d}p_{\lambda}(|\alpha_{0j}|), \label{pls}
\end{equation}
where $d$ is the dimension of $M_i$, and $p_{\lambda}(\cdot)$ is a
penalty function with a tuning parameter $\lambda$. In our
empirical analysis, the penalty function is taken to the SCAD
penalty proposed by \cite{fan2001variable}. The derivative of the
SCAD penalty is
$$
p_\lambda^{\prime}(|\beta|)=\lambda\left\{I(|\beta| \leq
\lambda)+\frac{(a \lambda-|\beta|)_{+}}{(a-1) \lambda}
I(|\beta|>\lambda)\right\},
$$
where $a=3.7$ as suggested in \cite{fan2001variable}. As in
\citet{guo2023highdim}, the tuning parameter $\lambda$ is selected
by the high-dimensional BIC (HBIC) criterion in
\cite{wang2013calibrating}. Indeed, we use the R-code developed by
\citet{guo2023highdim} for our empirical analysis in Section 4.

Minimizing (\ref{pls}) with respect to $\alpha_0$ and $\alpha_1$
results in their partially penalized least squares estimate
$\{\widehat{\alpha}_0, \widehat{\alpha}_1\}$. That is,
\begin{equation}
(\widehat{\alpha}_0, \widehat{\alpha}_1) = \underset{\alpha_0,
\alpha_1}{\arg\min} \left[
 \frac{1}{2n} \sum_{i=1}^n (Y_i-M_i^T\alpha_0-X_i^T\alpha_1)^2 +
\sum_{j=1}^{d}p_{\lambda}(|\alpha_{0j}|) \right].
\end{equation}
In (\ref{pls}), the penalty is applied only to $\alpha_0$, the
coefficient of financial metrics variables,  while the sector
coefficients, $\alpha_1$, are not penalized. Thus, all sector
indicators and active financial metric variables are retained in
the final selected model. In other words, the partially  penalized
least squares method enables less useful financial variables to be
removed while keeping all sector indicators in the model.

One might obtain least squares estimate $\widehat{\Gamma}$ of
$\Gamma$ based on model (\ref{eq3-4}), and then obtain an estimate
of $\beta$ by $\widehat{\Gamma}\hat{\alpha}_0$. This approach may
work well when $M$ is low dimensional, but it cannot control
cumulative errors in the presence of high-dimensional $M$
variables. \citet{guo2023highdim} advocated to estimate the
indirect effect by
\begin{equation}
\hat{\beta} = \hat{\gamma} - \hat{\alpha}_1,
\end{equation}
where $\hat{\gamma}$ is the least squares estimate of the total
effect model,
\[
Y_i=\gamma^TX_i+\eps^*_i.
\]


In practice, it is of interest to test whether there exists
indirect effect. This can be formulated as a statistical
hypothesis as follows:
\begin{equation}
H_0:\beta=0 \qquad \text{versus} \qquad H_1:\beta\neq0.
\label{indirect}
\end{equation}
\citet{guo2023highdim} developed a Wald test for (\ref{indirect}).
Furthermore, the authors developed an $F$-type test for
\begin{equation}
H_0:\alpha_1=0 \qquad \text{versus} \qquad H_1:\alpha_1\neq0
\label{direct}
\end{equation}
which can be used to examine whether direct sector effects are
significant or not. In our empirical analysis, we use the Wald
test for indirect effect and the F-type test for direct effect.


Since sector membership is not randomly assigned, these results are interpreted as relationships between sector, financial characteristics, and stock returns rather than as proof that tariff policy directly caused the differences.

\

\section{Real Data Analysis}
\label{sec:real-data-analysis}

In this section, we apply the high-dimensional mediation model
introduced in Section 3 for data collected over the five
tariff-policy windows, whose detailed dates are given in
Table~\ref{tab:results-event-windows}, which depicts the  average
log return over all stocks under consideration and approximate
return over each window. It is of great interest to study (a)
whether S\&P 500 firms reacted to the tariff episode, and (b) how
the reaction  changed as the policy environment moved from initial
implementation, to escalation, to temporary relief, to renewed
uncertainty, and finally to a longer adjustment period. This
sequencing is important because tariff news did not arrive as one
isolated shock. Instead, investors observed a series of policy
announcements, retaliations, partial reversals, and
sector-specific threats. Prior research shows that trade policy
uncertainty can reduce investment and economic activity even
before the full real effects of tariffs are observed in accounting
data \citep{handley_limao_2015,caldara_et_al_2020}. Tariffs can
also raise input and final-goods prices, alter supply chains, and
shift the incidence of trade policy toward domestic firms and
consumers \citep{amiti_redding_weinstein_2019}. The five-window
design therefore allows the empirical analysis to distinguish
between the initial tariff shock, the market collapse associated
with escalation, the one-day relief rally, the renewed decline
after uncertainty returned, and the longer-run repricing through
year-end.

\

\begin{xltabular}{\textwidth}{L{1.4cm} L{3.2cm} Y C{2.1cm} C{1.8cm}}
\caption{Event Windows and Market-Level Returns. Mean log return
is  the average cumulative log return across the stocks in the
sample during event window, and approx. return converts that mean
log return into a conventional percentage return using $\exp$(mean
log return)$-1$. }
\label{tab:results-event-windows} \\
\toprule
Scenario & Window & Main policy/event interpretation & Mean log return & Approx. return \\
\midrule
\endfirsthead

\multicolumn{5}{l}{\textit{Table \thetable{} continued}} \\
\toprule
Scen. & Window & Main policy/event interpretation & Mean log return & Approx. return \\
\midrule
\endhead

\bottomrule
\endlastfoot

S1 & Feb. 19 close\par --Mar. 13 close & Initial tariff-driven decline & -0.084 & -8.1\% \\
S2 & Mar. 25 close\par --Apr. 8 close & Escalation collapse & -0.142 & -13.2\% \\
S3 & Apr. 8 close\par --Apr. 9 close & Policy-shock relief jump & 0.080 & 8.3\% \\
S4 & Apr. 9 close\par --Apr. 21 close & Renewed uncertainty decline & -0.039 & -3.8\% \\
S5 & Feb. 19 close\par --Dec. 31 close & Long-run adjustment & 0.036 & 3.6\% \\
\end{xltabular}

\

\

\begingroup
\small
\setlength{\tabcolsep}{3pt}
\renewcommand{\arraystretch}{1.15}

\begin{xltabular}{\textwidth}{L{3.3cm} C{2.0cm} C{2.0cm} C{2.0cm} C{2.0cm} C{2.0cm}}
\caption{Significant direct sector effects across tariff-policy scenarios}
\label{tab:direct-sector-effects} \\
\toprule
Sector & S1 & S2 & S3 & S4 & S5 \\
\midrule
\endfirsthead

\multicolumn{6}{l}{\textit{Table \thetable{} continued}} \\
\toprule
Sector & S1 & S2 & S3 & S4 & S5 \\
\midrule
\endhead

\bottomrule
\multicolumn{6}{L{\textwidth}}{
\textit{Notes:} Entries are direct sector effects relative to Utilities, with standard errors in parentheses. Only effects with absolute test statistics of at least 1.96 are shown. Dashes indicate sector-scenario effects that are not reported as statistically significant under this rule. Utilities is the omitted reference sector. Full direct and indirect estimates are reported in the appendix.} \\
\endlastfoot

Basic Materials & -0.58 (0.22) & -1.12 (0.22) & 1.12 (0.22) & \textemdash & \textemdash \\

Communication Services & -0.79 (0.23) & -0.87 (0.23) & 0.95 (0.23) & -0.99 (0.24) & \textemdash \\

Consumer Cyclical & -0.73 (0.19) & -0.99 (0.19) & 0.96 (0.18) & -0.74 (0.19) & \textemdash \\

Consumer Defensive & \textemdash & \textemdash & \textemdash & \textemdash & -0.55 (0.22) \\

Energy & -0.51 (0.22) & -1.82 (0.22) & 1.04 (0.22) & -0.60 (0.23) & \textemdash \\

Financial Services & -0.49 (0.19) & -1.05 (0.18) & 0.76 (0.18) & -0.84 (0.18) & \textemdash \\

Healthcare & \textemdash & -0.58 (0.19) & \textemdash & -0.93 (0.18) & \textemdash \\

Industrials & -0.56 (0.17) & -0.99 (0.17) & 1.00 (0.17) & -0.81 (0.18) & \textemdash \\

Real Estate & \textemdash & -0.58 (0.21) & \textemdash & \textemdash & -0.50 (0.23) \\

Technology & -1.04 (0.17) & -1.54 (0.17) & 1.60 (0.17) & -1.12 (0.17) & \textemdash \\

\end{xltabular}

\endgroup

The complete direct and indirect sector effect estimates and
their standard errors are displayed in Table~\ref{tab:appendix-complete-direct-effects} and
\ref{tab:appendix-complete-indirect-effects}, respectively,
and complete selected financial-mediator estimates along with their
standard errors are depicted in
Table~\ref{tab:appendix-complete-selected-mediators}. Across the
five windows, the strongest and most consistent evidence appears
in the direct sector effects. See
Table~\ref{tab:direct-sector-effects} for the significant sector
effects. From Table~\ref{tab:appendix-complete-indirect-effects},
it seems that sector indirect effects on the responses are not
statistically distinguishable from zero, and therefore are not
emphasized as primary findings. The selected financial mediators
(see Table~\ref{tab:selected-mediators} for details) are used
instead to interpret firm-level channels within each window. This
distinction is important: sector effects may imply where the
market response was concentrated, while the financial mediators
describe the firm characteristics most associated with return
differences within each scenario.

\newpage

\begin{xltabular}{\textwidth}{L{1.2cm} Y L{3.0cm} C{2.7cm}}
\caption{Selected financial mediators by scenario}
\label{tab:selected-mediators} \\
\toprule
Scen. & Financial mediator & Transformation & Coefficient (SE) \\
\midrule
\endfirsthead

\multicolumn{4}{l}{\textit{Table \thetable{} continued}} \\
\toprule
Scen. & Financial mediator & Transformation & Coefficient (SE) \\
\midrule
\endhead

\bottomrule
\multicolumn{4}{L{\textwidth}}{\footnotesize \textit{Notes:} Table reports selected mediators with the largest substantive role in the scenario-level interpretation. Full selected-mediator outputs are reported in the appendix.} \\
\endlastfoot

S1 & Short-term investments & 3-year growth & -0.289 (0.007) \\
S1 & R\&D expense & QoQ change & -0.186 (0.007) \\
S1 & Capital expenditure & YoY change & 0.151 (0.004) \\
\ \\
S2 & EBIT & 3-year growth & 0.143 (0.004) \\
S2 & Capital expenditure & YoY change & 0.133 (0.003) \\
S2 & Revenue & QoQ change & 0.119 (0.003) \\
S2 & Operating cash flow & QoQ change & -0.099 (0.003) \\
\ \\
S3 & R\&D expense & QoQ change & 0.158 (0.003) \\
S3 & Short-term investments & 3-year growth & 0.140 (0.003) \\
S3 & Revenue & 3-year growth & 0.132 (0.002) \\
\ \\
S4 & Income before tax & YoY change & -0.208 (0.002) \\
S4 & Stock-based compensation & 3-year growth & -0.208 (0.002) \\
S4 & Cash and cash equivalents & YoY change & 0.139 (0.002) \\
\ \\
S5 & Goodwill & YoY change & -0.164 (0.012) \\
S5 & R\&D expense & QoQ change & -0.149 (0.018) \\
S5 & Free cash flow & YoY change & 0.111 (0.012) \\

\end{xltabular}

Table~\ref{tab:test} depicts the Wald test statistics for indirect
effects in (\ref{indirect}) and the $F$-type test statistics for
direct effects in (\ref{direct}) and their P-values. As shown in
Table~\ref{tab:test}, the impacts of sector membership on stock
return can be quite different over different periods. For example,
both direct effect and indirect effect of sector exposure are
significant over the initial tariff-driven decline period (i.e.
S1), but neither direct effect nor indirect effect is significant
over the long-run adjustment (i.e. S5). Table~\ref{tab:test}
indicates that the indirect effect of sector exposure is
significant at level 0.05 only for Scenario S1. This implies the
sector exposure has significant impact on the stock return through
financial metrics only during the initial period. This might be
interpreted as during the initial period, financial metrics for
most stock are highly associated with sector exposure, but after
the initial period, most firms in S\&P 500 got well prepared for
tariff, and the financial metrics become insensitive to sector
exposure. During the first four periods, the direct effects are
all very significant. This seems to be expected as the sector
exposure has impact on stock return when there are dramatic
changes on market. Both indirect effects and direct effects are
not significant for S5, which is the period from February 19, 2026
to December 31, 2026. That implies that the sector exposure does
not make difference on stock return during this period.

\begin{xltabular}{\textwidth}{C{1.4cm} | C{3.5cm} C{1.8cm} |  C{4.0cm} C{1.8cm}}
\caption{Wald Test Statistics for Indirect Effects in (\ref{indirect}) and $F$-type Test Statistics for Direct Effects in (\ref{direct}) and Their P-values}
\label{tab:test}\\
\toprule
Scenario & Wald test Statistics & P-value & $F$-type test Statistics & P-value \\
\midrule
\endhead

\bottomrule
\endlastfoot

S1 & 29.7030    &  0.0018  &  91.9949     &   <0.0001 \\
S2 &  8.1434    &  0.7004  & 222.9181     &   <0.0001 \\
S3 & 11.9844    &  0.3648  & 208.1513     &   <0.0001 \\
S4 & 10.5422    &  0.4824  & 134.6167     &   <0.0001 \\
S5 &  9.6870    &  0.5587  &  11.2421     &   0.42320  \\
\end{xltabular}

\subsection{Scenario 1: initial tariff-driven decline}

Scenario 1 captures the first broad equity-market decline
associated  with the 2025 tariff episode. The window begins at the
February 19 market peak and ends at the March 13 local low. It
covers the transition from tariff announcement to implementation
and sector expansion. On February 1, the administration announced
additional tariffs on imports from Canada, Mexico, and China,
including 25 percent tariffs on Canada and Mexico, a 10 percent
tariff on China, and a lower 10 percent rate on Canadian energy
\citep{whitehouse_tariffs_canada_mexico_china_2025}. On March 4,
these tariffs took effect and prompted retaliation from major
trading partners \citep{ap_tariffs_retaliation_2025}. On March 12,
expanded steel and aluminum tariff actions became effective,
widening the tariff episode from country-specific measures to
direct input-cost pressure on industrial supply chains
\citep{whitehouse_aluminum_2025, whitehouse_steel_2025}. The
average stock return in the frozen S\&P 500 universe was -0.084
log points, or approximately -8.1 percent.

The sector results indicate that rather than this initial decline was not a narrow response to one protected industry, it was a broad repricing of trade-policy risk. Technology shows the largest direct effect, with a coefficient of -1.04 relative to Utilities. This is the central result of Scenario 1. Technology firms are especially sensitive to tariff uncertainty because their valuations depend heavily on expected future cash flows, global supply chains, imported hardware components, and international demand. The selected financial mediators reinforce this interpretation. Growth in short-term investments has the largest selected mediator coefficient, -0.289, and recent R\&D growth has a coefficient of -0.186. These variables are closely associated with firms whose valuations depend on future-oriented investment rather than only current earnings. The result is consistent with the notion that when future rules over imports, costs, and cross-border production become less predictable, long-duration growth firms can be repriced sharply \citep{handley_limao_2015,caldara_et_al_2020}.

Communication Services has the second-largest negative direct effect in Scenario 1, with a coefficient of -0.79 relative to the Utilities baseline. The sector includes telecommunications, media, and digital-platform firms, many of which rely on globally sourced communications equipment and electronics. In March 2025, PwC estimated that total annual tariff measures affecting the broader Technology, Media, and Telecommunications industry could increase from approximately \$13 billion to \$139 billion \citep{pwc_tmt_tariff_analysis_2025}. This level of exposure is consistent with the relatively sharp repricing of Communication Services during the initial tariff-driven decline.

Consumer Cyclical has the third largest negative direct effect in Scenario 1. The sector has a direct effect of -0.73
relative to Utilities. This result is economically consistent with
the March implementation of tariffs on major U.S. trading
partners. Consumer Cyclical firms include companies exposed to
autos, retail supply chains, leisure, restaurants, and
discretionary consumption. Tariffs can affect these firms through
higher import costs, margin compression, and reduced household
purchasing power. \citet{amiti_redding_weinstein_2019} show that
tariff costs can pass through into domestic prices and affect
domestic buyers, which supports the interpretation that investors
were not only pricing foreign retaliation but also the possibility
of weaker consumer demand and higher costs within the United
States.

Industrials, Basic Materials, Energy, and Financial Services also have significant negative direct effects in Scenario 1, but they are best interpreted as supporting evidence rather than separate focal points. Industrials and Basic Materials connect naturally to steel, aluminum, transportation, and capital-goods supply chains. Energy's negative effect is notable because Canadian energy received a lower tariff rate, suggesting that broad demand expectations and policy uncertainty mattered alongside direct statutory tariff exposure. Financial Services' negative effect is consistent with a macro-financial repricing channel: as trade uncertainty rose, investors also marked down sectors sensitive to credit conditions, risk appetite, and expected economic growth. Overall, Scenario 1 shows that the first tariff decline penalized globally exposed and future-growth-oriented firms most heavily, rather than producing a simple split between protected and unprotected industries.

\subsection{Scenario 2: escalation collapse}

Scenario 2 captures the sharpest decline in the study. The window runs from the March 25 close through the April 8 close, and the average stock return was -0.142 log points, or approximately -13.2 percent. This period differs from Scenario 1 because tariff risk moved from initial implementation into rapid escalation. On March 25, the White House announced a 25 percent tariff on goods from countries importing Venezuelan oil \citep{whitehouse_venezuelan_oil_tariff_2025}. On March 26, the administration announced a 25 percent tariff on imported automobiles and certain automobile parts, with automobiles becoming subject to the tariff on April 3 \citep{whitehouse_auto_tariffs_2025}. On April 2, the administration announced reciprocal tariffs, including a 10 percent baseline tariff and higher country-specific tariff rates scheduled to begin April 9 \citep{whitehouse_reciprocal_tariffs_2025}. China responded on April 4 with 34 percent tariffs on U.S. goods and export restrictions on some rare earths \citep{reuters_china_retaliation_apr4_2025}. On April 7, the president threatened an additional 50 percent tariff on China if China did not withdraw its retaliation \citep{reuters_trump_additional_china_tariffs_apr7_2025}. Thus, Scenario 2 captures the market's reaction to a tariff shock that had become broader, more retaliatory, and more difficult to bound.

The largest direct effect in Scenario 2 is Energy, with a coefficient of -1.82 relative to Utilities. This distinguishes Scenario 2 from Scenario 1. The Venezuelan-oil-related tariff announcement directly connected trade policy to energy geopolitics, but the negative Energy response should not be interpreted simply as a direct tariff-cost effect. Energy stocks are also highly sensitive to expected global growth and commodity demand. A broader trade war raises recession risk, lowers expected industrial activity, and can pressure oil demand expectations. The large Energy coefficient suggests that investors viewed the escalation phase as both an economic slowdown and a trade policy shock.

Technology remains the second major focal sector, with its direct effect being -1.54. In this window, the China-specific escalation became much more central. China's retaliation included tariffs and rare-earth export curbs, while the U.S. threatened additional China tariffs above the reciprocal-tariff framework \citep{reuters_china_retaliation_apr4_2025,reuters_trump_additional_china_tariffs_apr7_2025}. Technology firms are especially exposed to semiconductor inputs, electronics supply chains, global manufacturing networks, and China-linked demand. The April 2 reciprocal tariff order also excluded some goods because they might become subject to future sector-specific actions, including semiconductors, which likely preserved uncertainty rather than eliminating it \citep{whitehouse_reciprocal_tariffs_2025}. The selected mediators support a distinction between durable operating strength and immediate cash-flow stress. EBIT three-year growth has a positive selected coefficient of 0.143, capital expenditure growth has a coefficient of 0.133, and recent revenue growth has a coefficient of 0.119. These positive coefficients suggest that firms with stronger operating or investment capacity were relatively more resilient inside the collapse. At the same time, operating cash-flow growth has a negative selected coefficient of -0.099, indicating that recent cash-flow dynamics were not uniformly protective. In the escalation phase, the market appears to have differentiated between stable longer-run operating strength and short-run cash-flow exposure.

The remaining significant sectors support the interpretation of a broad escalation collapse. Basic Materials has a direct effect of -1.12, Financial Services -1.05, Consumer Cyclical -0.99, and Industrials -0.99. These are economically large effects and align with the policy reasoning that reciprocal tariffs, auto tariffs, steel and aluminum actions, and Chinese retaliation together affected input costs, downstream demand, capital goods demand, and financing conditions. Healthcare and Real Estate also become significantly negative in Scenario 2, though with smaller coefficients. Their inclusion suggests that by early April the tariff episode was no longer confined to goods-producing or globally integrated sectors. It had become a market-wide stress event.

\subsection{Scenario 3: policy-shock relief jump}

Scenario 3 is the shortest window but one of the most informative. It measures the close-to-close return from April 8 to April 9, the day on which the administration suspended many country-specific reciprocal tariffs for 90 days while increasing tariffs on China. The average stock return was 0.080 log points, or approximately 8.3 percent. The White House order suspended enforcement of many country-specific reciprocal rates until July 9, while keeping a 10 percent baseline tariff and raising the China-specific rate to 125 percent \citep{whitehouse_reciprocal_pause_china_escalation_2025}. Reuters reported that the reversal sent U.S. stocks sharply higher, reflecting relief that the broadest tariff rates would not immediately apply to most trading partners \citep{reuters_apr9_relief_rally_2025}. This window therefore captures a relief rally, not a full resolution of the trade war.

The strongest direct effect is again Technology, but the sign
reverses sharply. Technology has a direct effect of 1.60 relative
to Utilities. This reversal is central to the paper's
interpretation. The same sector that was most penalized during the
initial decline and escalation collapse was also the most
positively revalued when the worst-case global tariff path was
temporarily suspended. This pattern is consistent with the effects
of changes in discount rates and policy uncertainty. When tariff
uncertainty rises, long-duration growth firms are punished. When
the immediate possibility of a broad global tariff regime falls,
those same firms rebound sharply. The selected mediators
strengthen this interpretation. Recent R\&D growth has a positive
coefficient of 0.158, short-term investment growth has a
coefficient of 0.140, and long-run revenue growth has a
coefficient of 0.132. These are the same types of future-oriented
variables that were penalized in the first decline, but they
became positively associated with returns during the relief jump.

The second major example is the group of cyclical and trade-exposed sectors: Basic Materials, Energy, Industrials, and Consumer Cyclical. Basic Materials has a direct effect of 1.12, Energy 1.04, Industrials 1.00, and Consumer Cyclical 0.96. Their positive effects indicate that the April 9 rally was not limited to Technology. It also reflected a repricing of recession and supply-chain risk. Materials and industrial firms benefit when the market lowers the probability of an immediate collapse in global trade volumes. Consumer Cyclical firms benefit when investors reduce the expected hit to household purchasing power, autos, retail margins, and imported goods costs. Energy's positive effect is also consistent with an improved demand outlook after the tariff pause.

However, Scenario 3 should not be interpreted as evidence that tariff risk disappeared. The April 9 order preserved a 10 percent baseline tariff and escalated China-specific duties. China subsequently raised its own duties on U.S. goods to 125 percent \citep{reuters_china_125_tariffs_apr11_2025}. The one-day jump is therefore best interpreted as a relief from the most extreme multi-country tariff scenario, not as a resolution of the underlying trade conflict. This distinction helps explain why the market rebound was sharp but temporary, and why renewed uncertainty appears in Scenario 4.

\subsection{Scenario 4: renewed uncertainty decline}

Scenario 4 covers the decline from the April 9 close through the April 21 close. The average stock return was -0.039 log points, or approximately -3.8 percent. This window differs from Scenario 2 because the broad reciprocal-tariff shock had already been partially paused. The negative return instead reflects the return of unresolved uncertainty. China raised duties on U.S. goods to 125 percent after the U.S. maintained intense pressure on China \citep{reuters_china_125_tariffs_apr11_2025}. On April 14, Reuters reported that the administration had initiated Section 232 investigations into pharmaceutical and semiconductor imports, setting the stage for possible sector-specific tariffs \citep{reuters_section232_pharma_semiconductors_2025}. On April 21, U.S. equities sold off after renewed presidential criticism of Federal Reserve Chair Jerome Powell raised concerns about central-bank independence, while markets were still processing tariff uncertainty \citep{reuters_powell_selloff_apr21_2025}. Scenario 4 therefore combines trade escalation, sector-specific tariff threats, and monetary-policy credibility concerns.

Technology remains significantly negative, with a direct effect of -1.12. The Technology result in Scenario 4 differs from the Technology result in Scenario 2. In Scenario 2, the shock was broad reciprocal-tariff escalation. In Scenario 4, the market was responding more specifically to unresolved China tensions and possible semiconductor tariffs. The April 14 Section 232 investigation into semiconductor imports created uncertainty over a sector that is central to Technology earnings and supply chains. The persistence of Technology's negative direct effect after the April 9 relief rally shows that investors did not treat the tariff pause as a complete removal of technology-sector risk.

Healthcare becomes one of the most important Scenario 4 results. Its direct effect is -0.93 relative to Utilities. Healthcare was not a central significant sector in Scenario 1, but it becomes significant after pharmaceutical imports entered the tariff discussion. This is precisely why the scenario-by-scenario structure matters. A generic ``tariffs hurt exposed firms'' explanation would miss the timing. In Scenario 4, the relevant policy news included possible pharmaceutical tariffs, which directly changed the risk profile of drug developers, healthcare suppliers, and firms dependent on globally distributed production networks. The selected mediators also point toward profitability and balance-sheet sensitivity. Income before tax year-over-year change has a coefficient of -0.208, and stock-based compensation three-year growth has a coefficient of -0.208. Cash and cash equivalents year-over-year change has a positive coefficient of 0.139. This pattern suggests that during renewed uncertainty, the market rewarded liquidity buffers while penalizing firms with earnings or compensation structures that could become more vulnerable under higher uncertainty and tighter financial conditions.

Financial Services is also significant in Scenario 4, with a direct effect of -0.84. This result is especially tied to the April 21 event. Financial firms are not directly tariffed in the same way as imported goods, but they are highly exposed to interest-rate expectations, credit conditions, and institutional confidence. The market reaction to criticism of the Federal Reserve occurred while tariff uncertainty was still unresolved. Therefore, the Financial Services result should be interpreted as a macro-financial channel: tariff uncertainty and monetary-policy uncertainty reinforced one another. In this window, the tariff episode had moved beyond import costs alone and had become connected to broader concerns about policy credibility, discount rates, and financial stability.

\subsection{Scenario 5: long-run adjustment}

Scenario 5 covers the long window from the February 19 close through the December 31 close. The average stock return was 0.036 log points, or approximately 3.6 percent. This window measures the broader adjustment from the initial tariff shock, the April collapse, the relief rally, to later stabilization. On May 12, the United States and China announced a Geneva agreement that suspended major portions of the tariff escalation for 90 days and established a mechanism for continued talks \citep{whitehouse_us_china_geneva_statement_2025}. Reuters reported that by mid-May the S\&P 500 had rebounded sharply from its April low, helped by the U.S.-China trade truce \citep{reuters_may_rebound_tariff_truce_2025}. By June 27, the S\&P 500 and Nasdaq had returned to record highs, with easing tariff fears and earnings strength cited as major drivers \citep{reuters_sp500_record_jun27_2025}. By December 31, markets approached the end of a volatile but ultimately positive year, marked by both tariff uncertainty and strong enthusiasm around artificial intelligence \citep{reuters_year_end_market_dec31_2025}.

The long-window results differ sharply from the short-window results. Technology, Energy, Industrials, and Consumer Cyclical are no longer significantly negative relative to Utilities. Instead, only Consumer Defensive and Real Estate show significant negative direct effects. Consumer Defensive has a direct effect of -0.55, and Real Estate has a direct effect of -0.50. This means that these sectors underperformed the Utilities reference category after the year's recovery and repricing. The long-window pattern therefore suggests that the market eventually moved from panic over broad tariff escalation toward a selective adjustment based on sector defensiveness, interest-rate sensitivity, cash-flow durability, and participation in the recovery.

Real Estate is the first important long-window example. Its significant negative direct effect is consistent with the sector's sensitivity to discount rates, financing conditions, and economic uncertainty. Even after tariff fears eased, the year still contained substantial uncertainty over inflation, interest rates, and policy credibility. Real Estate firms depend heavily on financing costs and long-duration asset valuations. A recovery led by earnings strength, optimism related to the growing AI sector, and relief from tariff panic does not necessarily benefit Real Estate in the same way it benefits growth or cyclical sectors.

Consumer Defensive is the second important example. The sector's negative long-window effect should be interpreted as relative underperformance in a risk-on recovery rather than as a direct tariff-collapse story. Defensive firms often hold up better during acute selloffs, but they can lag when markets rebound and investors rotate back into growth, cyclicals, and financials. The negative long-window coefficient is therefore consistent with a rotation away from defensive exposures as tariff fears eased and broader equity-market risk appetite recovered.

The selected financial mediators in Scenario 5 also differ from the short-run windows. Goodwill year-over-year change has a coefficient of -0.164, R\&D expense quarter-over-quarter change has a coefficient of -0.149, and free cash flow year-over-year change has a positive coefficient of 0.111. These results suggest that the long-run adjustment was less about immediate tariff exposure and more about balance-sheet quality, expense discipline, and cash generation. This is consistent with asset-pricing evidence that profitability and investment patterns help explain cross-sectional stock returns \citep{fama_french_2015}. It is also consistent with accounting-based return research showing that cash-flow and earnings information are central to firm-level return variation \citep{callen_segal_2004}. In short, Scenario 5 shows that once the acute tariff panic faded, investors returned to fundamentals: firms with stronger cash generation and cleaner balance-sheet signals were treated more favorably, while firms with more fragile intangible, spending, or defensive profiles lagged.

\section{Conclusion}

This paper studies how S\&P 500 firms reacted to the 2025 tariff-policy episode across five event windows. The results show that tariff news did not produce one uniform market response. Sector effects continued shifting as the policy environment moved from initial tariff implementation, to escalation, to temporary relief, to resurfaced uncertainty, and then finally to long-term adjustment.
The clearest findings are in the direct sector effects. Technology was the most negatively affected sector during the initial decline and escalation collapse, but it also rebounded most strongly during the April 9 relief rally. Energy was also one of the most negatively affected during the escalation-collapse window, while Healthcare became important during the resurfaced uncertainty window after pharmaceutical-import concerns emerged. These differences show why the tariff episode must be studied by scenario rather than as one full-period return.

The selected financial mediators suggest that investors also differentiated firms by financial structure. Short-term reactions were associated more with growth, investment, liquidity, profitability, and cash-flow variables, while the long-term window valued balance-sheet quality, expense discipline, and free cash flow.

Overall, the analysis shows that tariff-policy shocks can create rapidly changing differences in stock-market performance across firms. The market response depends not only on whether tariffs rise or fall, but also on the specific countries, sectors, and uncertainties involved at each point in time. Future work could extend the analysis using firm-level import exposure, supply-chain data, or international revenue shares to measure tariff exposure more directly.

\appendix

\section{Appendix}
\label{app:first}

\setcounter{table}{0}
\renewcommand{\thetable}{A\arabic{table}}
\renewcommand{\theHtable}{A.\arabic{table}}

\subsection{Variable Dictionary and Data Construction}

The public code repository provides a complete variable dictionary for the identifier, 11 sector indicators, 215 financial features, four reference-date fields, and five scenario-specific outcome variables at \href{https://github.com/qilankhong/sp500-tariff-mediation-2025/blob/main/docs/Appendix_Variable_Dictionary.csv}{\texttt{docs/Appendix\_Variable\_Dictionary.csv}}. The file records the paper variable name, the variable role, the data source, the definition, the construction rule, the original unit, and data pre-processing notes.

\begingroup
\small
\setlength{\tabcolsep}{5pt}
\renewcommand{\arraystretch}{1.15}

\begin{xltabular}{\textwidth}{L{7.0cm} C{3.0cm} Y}
\caption{Dataset construction summary}
\label{tab:appendix-dataset-summary} \\
\toprule
Item & Count & Notes \\
\midrule
\endfirsthead

\multicolumn{3}{l}{\textit{Table \thetable{} continued}} \\
\toprule
Item & Count & Notes \\
\midrule
\endhead

\bottomrule
\endlastfoot

Stocks & 503 & Frozen S\&P 500 stock-level analysis universe \\
Daily stock price & 252 trading days & FMP dividend-adjusted daily stock prices from January 2 through December 31, 2025 \\
Sector indicators & 11 & FMP company-profile sector field; Utilities is the omitted reference sector in the model \\
Raw income-statement variables & 17 & Quarterly FMP income-statement fields \\
Raw balance-sheet variables & 22 & Quarterly FMP balance-sheet fields \\
Raw cash-flow variables & 4 & Quarterly FMP cash-flow-statement fields \\
Candidate financial mediators & 215 & Forty-three raw accounting series expanded into five candidate features each \\
Usable financial mediators after screening & 214 & One all-missing mediator was removed before estimation \\
Event-window outcomes & 5 & Five close-to-close tariff-policy stock-return windows \\

\end{xltabular}

\endgroup

\subsection{Sample Coverage}

\begingroup
\small
\setlength{\tabcolsep}{5pt}
\renewcommand{\arraystretch}{1.15}

\begin{xltabular}{\textwidth}{L{5.0cm} C{3.0cm} Y}
\caption{Sector counts in the frozen analysis universe}
\label{tab:appendix-sector-counts} \\
\toprule
Sector & Stock observations & Notes \\
\midrule
\endfirsthead

\multicolumn{3}{l}{\textit{Table \thetable{} continued}} \\
\toprule
Sector & Stock observations & Notes \\
\midrule
\endhead

\bottomrule
\endlastfoot

Basic Materials & 23 & Included as a sector indicator \\
Communication Services & 25 & Included as a sector indicator \\
Consumer Cyclical & 53 & Included as a sector indicator \\
Consumer Defensive & 37 & Included as a sector indicator \\
Energy & 23 & Included as a sector indicator \\
Financial Services & 70 & Included as a sector indicator \\
Healthcare & 60 & Included as a sector indicator \\
Industrials & 72 & Included as a sector indicator \\
Real Estate & 30 & Included as a sector indicator \\
Technology & 78 & Included as a sector indicator \\
Utilities & 32 & Omitted reference sector \\

\end{xltabular}

\endgroup

\begingroup
\small
\setlength{\tabcolsep}{4pt}
\renewcommand{\arraystretch}{1.15}

\begin{xltabular}{\textwidth}{L{1.2cm} L{3.4cm} C{2.0cm} C{2.4cm} Y}
\caption{Event-window return coverage}
\label{tab:appendix-return-coverage} \\
\toprule
Scen. & Outcome & Return rows in complete window & Complete ticker observations & Coverage note \\
\midrule
\endfirsthead

\multicolumn{5}{l}{\textit{Table \thetable{} continued}} \\
\toprule
Scen. & Outcome & Return rows in complete window & Complete ticker observations & Coverage note \\
\midrule
\endhead

\bottomrule
\endlastfoot

S1 & \texttt{S1\_Decline} & 16 & 503 & Complete for all ticker observations \\
S2 & \texttt{S2\_Escalation\_Collapse} & 10 & 503 & Complete for all ticker observations \\
S3 & \texttt{S3\_Policy\_Shock\_Jump} & 1 & 503 & Complete for all ticker observations \\
S4 & \texttt{S4\_Uncertainty\_Decline} & 7 & 503 & Complete for all ticker observations \\
S5 & \texttt{S5\_Long\_Term\_Adjustment} & 218 & 500 & FI, IPG, and K have shorter FMP price histories in this window, with 202, 195, and 204 return rows, respectively \\

\addlinespace
\multicolumn{5}{L{\textwidth}}{\footnotesize \textit{Notes:} Return rows are daily log-return rows. Because each daily return is labeled by its ending trading date, a close-to-close window from start date to end date includes returns with start date $< t \leq$ end date.} \\

\end{xltabular}

\endgroup

\

\subsection{Complete Sector Effect Estimates}

\begingroup
\scriptsize
\setlength{\tabcolsep}{3pt}
\renewcommand{\arraystretch}{1.15}

\begin{xltabular}{\textwidth}{L{3.1cm} C{2.15cm} C{2.15cm} C{2.15cm} C{2.15cm} C{2.15cm}}
\caption{Complete direct sector effect estimates}
\label{tab:appendix-complete-direct-effects} \\
\toprule
Sector & S1 & S2 & S3 & S4 & S5 \\
\midrule
\endfirsthead

\multicolumn{6}{l}{\textit{Table \thetable{} continued}} \\
\toprule
Sector & S1 & S2 & S3 & S4 & S5 \\
\midrule
\endhead

\bottomrule
\endlastfoot

Basic Materials & -0.58 (0.22) & -1.12 (0.22) & 1.12 (0.22) & -0.29 (0.23) & -0.28 (0.25) \\
Communication Services & -0.79 (0.23) & -0.87 (0.23) & 0.95 (0.23) & -0.99 (0.24) & -0.41 (0.24) \\
Consumer Cyclical & -0.73 (0.19) & -0.99 (0.19) & 0.96 (0.18) & -0.74 (0.19) & -0.22 (0.21) \\
Consumer Defensive & 0.10 (0.20) & 0.30 (0.20) & -0.07 (0.20) & 0.25 (0.20) & -0.55 (0.22) \\
Energy & -0.51 (0.22) & -1.82 (0.22) & 1.04 (0.22) & -0.60 (0.23) & -0.38 (0.25) \\
Financial Services & -0.49 (0.19) & -1.05 (0.18) & 0.76 (0.18) & -0.84 (0.18) & -0.16 (0.20) \\
Healthcare & -0.15 (0.18) & -0.58 (0.19) & 0.32 (0.18) & -0.93 (0.18) & -0.28 (0.20) \\
Industrials & -0.56 (0.17) & -0.99 (0.17) & 1.00 (0.17) & -0.81 (0.18) & -0.15 (0.19) \\
Real Estate & -0.15 (0.21) & -0.58 (0.21) & 0.28 (0.20) & 0.11 (0.21) & -0.50 (0.23) \\
Technology & -1.04 (0.17) & -1.54 (0.17) & 1.60 (0.17) & -1.12 (0.17) & -0.22 (0.19) \\

\addlinespace
\multicolumn{6}{L{\textwidth}}{\footnotesize \textit{Notes:} Entries report direct sector effect estimates, with standard errors in parentheses. Utilities is omitted from the regression and is the reference sector. These are the complete sector estimates; the main text reports only the statistically emphasized direct effects.} \\

\end{xltabular}

\endgroup

\newpage

\begingroup
\scriptsize
\setlength{\tabcolsep}{3pt}
\renewcommand{\arraystretch}{1.15}

\begin{xltabular}{\textwidth}{L{3.1cm} C{2.15cm} C{2.15cm} C{2.15cm} C{2.15cm} C{2.15cm}}
\caption{Complete indirect sector effect estimates}
\label{tab:appendix-complete-indirect-effects} \\
\toprule
Sector & S1 & S2 & S3 & S4 & S5 \\
\midrule
\endfirsthead

\multicolumn{6}{l}{\textit{Table \thetable{} continued}} \\
\toprule
Sector & S1 & S2 & S3 & S4 & S5 \\
\midrule
\endhead

\bottomrule
\endlastfoot

Basic Materials & 0.16 (0.13) & -0.20 (0.11) & 0.03 (0.11) & -0.12 (0.11) & -0.16 (0.13) \\
Communication Services & 0.10 (0.16) & -0.06 (0.13) & 0.05 (0.14) & 0.02 (0.14) & 0.18 (0.13) \\
Consumer Cyclical & -0.09 (0.12) & -0.06 (0.09) & 0.14 (0.10) & -0.15 (0.09) & -0.15 (0.11) \\
Consumer Defensive & 0.17 (0.14) & -0.07 (0.11) & 0.02 (0.11) & 0.05 (0.09) & 0.02 (0.12) \\
Energy & -0.02 (0.13) & 0.00 (0.10) & 0.09 (0.11) & -0.09 (0.10) & -0.09 (0.13) \\
Financial Services & -0.17 (0.13) & -0.05 (0.09) & 0.16 (0.10) & -0.01 (0.08) & 0.03 (0.11) \\
Healthcare & 0.12 (0.12) & 0.03 (0.10) & -0.01 (0.10) & -0.07 (0.09) & -0.02 (0.11) \\
Industrials & 0.09 (0.11) & -0.05 (0.09) & 0.04 (0.09) & -0.00 (0.08) & -0.01 (0.10) \\
Real Estate & 0.07 (0.13) & -0.06 (0.10) & 0.16 (0.11) & -0.02 (0.10) & -0.11 (0.12) \\
Technology & -0.20 (0.11) & -0.04 (0.09) & 0.15 (0.09) & -0.10 (0.08) & -0.00 (0.10) \\

\addlinespace
\multicolumn{6}{L{\textwidth}}{\footnotesize \textit{Notes:} Entries report indirect sector effect estimates, with standard errors in parentheses. Utilities is the omitted reference sector. Individual sector-level indirect effects are not emphasized as primary findings in the main text because they are not statistically distinguishable from zero in the saved model outputs.} \\

\end{xltabular}

\endgroup

\subsection{Complete Selected-Mediator Estimates}

\begingroup
\scriptsize
\setlength{\tabcolsep}{4pt}
\renewcommand{\arraystretch}{1.12}

\begin{xltabular}{\textwidth}{L{1.1cm} Y C{2.4cm}}
\caption{Complete selected financial-mediator estimates}
\label{tab:appendix-complete-selected-mediators} \\
\toprule
Scen. & Selected mediator & Coefficient (SE) \\
\midrule
\endfirsthead

\multicolumn{3}{l}{\textit{Table \thetable{} continued}} \\
\toprule
Scen. & Selected mediator & Coefficient (SE) \\
\midrule
\endhead

\bottomrule
\endlastfoot

S1 & \path{IS:revenue_QoQ} & 0.076 (0.005) \\
S1 & \path{IS:revenue_CAGR3} & -0.102 (0.005) \\
S1 & \path{IS:grossProfit_YoY} & -0.126 (0.004) \\
S1 & \path{IS:grossProfit_QoQ} & 0.063 (0.004) \\
S1 & \path{IS:researchAndDevelopmentExpenses_QoQ} & -0.186 (0.007) \\
S1 & \path{IS:generalAndAdministrativeExpenses} & -0.063 (0.004) \\
S1 & \path{IS:interestExpense_CAGR3} & -0.082 (0.004) \\
S1 & \path{IS:ebitda_CAGR3} & -0.147 (0.004) \\
S1 & \path{IS:eps} & 0.089 (0.004) \\
S1 & \path{BS:shortTermInvestments_CAGR3} & -0.289 (0.007) \\
S1 & \path{BS:intangibleAssets} & 0.132 (0.004) \\
S1 & \path{BS:accountPayables_Avg5} & -0.083 (0.004) \\
S1 & \path{BS:retainedEarnings_YoY} & -0.137 (0.004) \\
S1 & \path{CF:capitalExpenditure_YoY} & 0.151 (0.004) \\
S1 & \path{CF:stockBasedCompensation_CAGR3} & -0.097 (0.004) \\
\addlinespace
S2 & \path{IS:revenue_QoQ} & 0.119 (0.003) \\
S2 & \path{IS:costOfRevenue} & 0.071 (0.004) \\
S2 & \path{IS:grossProfit_YoY} & -0.045 (0.004) \\
S2 & \path{IS:operatingIncome_YoY} & -0.092 (0.003) \\
S2 & \path{IS:interestExpense_YoY} & 0.098 (0.003) \\
S2 & \path{IS:incomeBeforeTax_YoY} & 0.056 (0.003) \\
S2 & \path{IS:ebit_CAGR3} & 0.143 (0.004) \\
S2 & \path{IS:eps} & 0.081 (0.003) \\
S2 & \path{BS:goodwill} & 0.024 (0.004) \\
S2 & \path{BS:intangibleAssets} & 0.067 (0.004) \\
S2 & \path{CF:netCashProvidedByOperatingActivities_QoQ} & -0.099 (0.003) \\
S2 & \path{CF:capitalExpenditure_YoY} & 0.133 (0.003) \\
S2 & \path{CF:freeCashFlow_QoQ} & -0.080 (0.003) \\
\addlinespace
S3 & \path{IS:revenue_QoQ} & -0.138 (0.002) \\
S3 & \path{IS:revenue_CAGR3} & 0.132 (0.002) \\
S3 & \path{IS:grossProfit_YoY} & 0.043 (0.002) \\
S3 & \path{IS:researchAndDevelopmentExpenses_QoQ} & 0.158 (0.003) \\
S3 & \path{IS:operatingIncome_YoY} & 0.100 (0.002) \\
S3 & \path{IS:ebit_CAGR3} & -0.131 (0.002) \\
S3 & \path{IS:eps} & -0.078 (0.002) \\
S3 & \path{BS:shortTermInvestments_CAGR3} & 0.140 (0.003) \\
S3 & \path{BS:propertyPlantEquipmentNet_CAGR3} & 0.086 (0.002) \\
S3 & \path{BS:intangibleAssets_Avg5} & -0.113 (0.002) \\
S3 & \path{CF:netCashProvidedByOperatingActivities_QoQ} & 0.054 (0.002) \\
S3 & \path{CF:capitalExpenditure_YoY} & -0.120 (0.002) \\
S3 & \path{CF:freeCashFlow_QoQ} & 0.090 (0.002) \\
\addlinespace
S4 & \path{IS:incomeBeforeTax_YoY} & -0.208 (0.002) \\
S4 & \path{BS:cashAndCashEquivalents_YoY} & 0.139 (0.002) \\
S4 & \path{BS:otherCurrentAssets_CAGR3} & -0.125 (0.002) \\
S4 & \path{BS:intangibleAssets_Avg5} & 0.084 (0.002) \\
S4 & \path{BS:deferredRevenue_CAGR3} & -0.159 (0.003) \\
S4 & \path{BS:commonStock_CAGR3} & -0.079 (0.002) \\
S4 & \path{CF:netCashProvidedByOperatingActivities_QoQ} & -0.081 (0.002) \\
S4 & \path{CF:stockBasedCompensation} & -0.069 (0.002) \\
S4 & \path{CF:stockBasedCompensation_CAGR3} & -0.208 (0.002) \\
\addlinespace
S5 & \path{IS:researchAndDevelopmentExpenses_QoQ} & -0.149 (0.018) \\
S5 & \path{IS:sellingAndMarketingExpenses_QoQ} & -0.143 (0.021) \\
S5 & \path{IS:operatingIncome_YoY} & 0.097 (0.011) \\
S5 & \path{IS:interestExpense_YoY} & -0.068 (0.011) \\
S5 & \path{IS:incomeBeforeTax_YoY} & -0.095 (0.010) \\
S5 & \path{IS:incomeTaxExpense_YoY} & -0.113 (0.011) \\
S5 & \path{IS:incomeTaxExpense_CAGR3} & 0.151 (0.012) \\
S5 & \path{IS:weightedAverageShsOutDil} & 0.095 (0.014) \\
S5 & \path{BS:netReceivables_Avg5} & 0.063 (0.012) \\
S5 & \path{BS:inventory} & 0.087 (0.011) \\
S5 & \path{BS:goodwill_YoY} & -0.164 (0.011) \\
S5 & \path{BS:shortTermDebt_YoY} & -0.112 (0.012) \\
S5 & \path{BS:shortTermDebt_CAGR3} & 0.095 (0.012) \\
S5 & \path{BS:commonStock} & 0.069 (0.012) \\
S5 & \path{BS:commonStock_CAGR3} & -0.115 (0.012) \\
S5 & \path{BS:totalStockholdersEquity_CAGR3} & -0.054 (0.011) \\
S5 & \path{CF:netCashProvidedByOperatingActivities_YoY} & 0.093 (0.011) \\
S5 & \path{CF:stockBasedCompensation_CAGR3} & -0.122 (0.012) \\
S5 & \path{CF:freeCashFlow_YoY} & 0.111 (0.011) \\

\addlinespace
\multicolumn{3}{L{\textwidth}}{\footnotesize \textit{Notes:} Entries report financial mediators selected in the saved model outputs, with standard errors in parentheses. Variable names follow the public variable dictionary. Prefixes indicate the FMP statement source: IS for income statement, BS for balance sheet, and CF for cash-flow statement.} \\

\end{xltabular}

\endgroup

\section{Code and Data Availability}

All analysis code is available in the public GitHub repository at \url{https://github.com/qilankhong/sp500-tariff-mediation-2025}. Private API keys, local cache files, and downloaded FMP-derived datasets are not included in the public repository. The repository provides the scripts needed to rebuild the analysis files with authorized FMP access, along with documentation for the computational workflow and validation checks.

\bibliographystyle{apalike} 

\bibliography{reference} 

\end{document}